# Exact solutions of macroscopic self-consistent electromagnetic fields and microscopic distribution of Vlasov-Maxwell system.

H. Lin

*State Key Laboratory of High Field Laser Physics, Shanghai Institute of*

*Optics and Fine Mechanics,*

*P. O. Box 800-211, Shanghai 201800, China;*

*linhai@siom.ac.cn*

()

Strict mathematics reveals that the strict solution of a Vlasov-Maxwell equation set cannot be of a zero-temperature mathematical form $f(r,v,t) = \int f(r,v,t) d^3 v \delta * (v - u(r,t))$. This universal property of Vlasov-Maxwell system can lead to a closed equation set of three macroscopic quantities: self-consistent fields $E$, $B$ and fluid velocity $u$, and hence their exact solutions. Strict solution of microscoipc distribution governed by self-consistent electromagnetic fields is found to be a universal extension of well-known BGK mode, which corresponds to $B \equiv 0$ case.

PACS: 41.75.-i, 29.27.Bd, 29.27.Fh, 52.27.Jt, 52.35.-g, 52.35.Fp, 52.35.Mw, 52.35.Sb, 52.65.-y.

Vlasov-Maxwell (V-M) equation set is the acknowledged theoretical basis of plasmas and beam physics[1]. The difficulty of strictly solving the V-M system is also evident. Few strict analytic solutions of microscopic distribution $f$, such as BGK mode [2], are found in the electrostatic case in which self-consistent magnetic field meets $B \equiv 0$. In more general cases with electromagnetic self-consistent field, strict analytic solution of $f$ is not yet found by now even though people have studied intensively a class of approximated solutions, which meet stationary condition $\partial_t f = 0$ [3-6].

On the other hand, even if we do not seek for exact microscopic distribution, it is still diffficult to obtain exact macroscopic self-consistent fields. It is well-known that standard procedure leads to macroscopic fluid equations being entangled with thermal pressure, which is associated with microscopic distribution function [1]. This causes different approximation treatment on the macroscopic fluid equations and hence self-consistent fields. For example, for plasma electrostatic wave, cold fluid theory [7-10] and warm fluid theory [11-15] can lead to different macroscopic description.

The importance of exact microscopic and macroscopic information obtained from the V-M system is self-evident. Many applications [16] demand microscopic and macroscopic information to be as exact as possible. By now, few progress in the theory of the V-M system [17] promote people to turn their attention to various high-efficiency numerical schemes on the V-M system [18-20].

For the V-M equation set (where $r$, $v$ and $t$ are independent variables)

$$\left[\partial_t + v \cdot \nabla - e\left[E + v \times B\right] \cdot \partial_{p(v)}\right] f = 0 \qquad (1)$$

$$\partial_t E = enu + \nabla \times B; \qquad (2)$$

$$\nabla \cdot E = -en + ZeN_i; \qquad (3)$$

$$\nabla \times E = -\partial_t B; \qquad (4)$$

$$\nabla \cdot B = 0. \qquad (5)$$

we have been also familiar with a standard procedure of deriving a macroscopic fluid motion equation

$$\partial_t u + \frac{\int (v-u)\nabla_r[vf]d^3v}{n} + \frac{e}{m}\frac{\int \left[E(r,t) + v \times B(r,t)\right] \cdot [\sqrt{1-v^2}]^3 * f d^3v}{n} = 0 \qquad (6)$$

where $p(v) = \frac{mv}{\sqrt{1-v^2/c^2}}$, $u = \frac{\int v f d^3v}{\int f d^3v}$, $n = \int f d^3v$, $e$ and $m$ are electronic charge and mass. According to textbooks [1], if following similar procedure, we can obtain coupled fluid equations in infinite number and hence find that this set is unsolvable unless introducing truncation approximation. Therefore, Eq.(1) and Eq(6) are not transparent enough because what they reflect is an entangled relation among $u$, $(E,B)$ and $pressure = \int(v-u)\nabla_r[(v-u)f]d^3v$. Eqs.(2-6) is thus unsolvable.

We should note that if self-consistent fields $(E,B)$ exist, the Vlasov equation (VE), or Eq.(1), cannot have a zero-temperature type solution. Namely, the microscopic distribution of charged particles cannot be of a general form $f(r,v,t) = \alpha(r,t)\delta[v-\beta(r,t)]$. This can be easily verified by inserting $\alpha(r,t)\delta[v-\beta(r,t)]$ into a VE

$$[\partial_t + v \cdot \nabla_r - e[E(r,t) + v \times B(r,t)] \cdot \partial_p][\alpha * \delta(v-\beta)]$$
$$= [\partial_t + v \cdot \nabla_r]\alpha * \delta$$
$$+ \alpha * \{[\partial_t + v \cdot \nabla_r](-\beta) * \delta' - e[E(r,t) + v \times B(r,t)] \cdot (\partial_p v) * \delta'\} \quad (7)$$

where $\delta'$ stands for the derivative of the Dirac function $\delta$ with respect to its variable and the relations $x\delta'(x) = -\delta(x)$ and $x\delta(x) = 0$ have been applied in above formula. Obviously, Eq.(7) is not definitely a VE of $\alpha * \delta(v-\beta)$. This fact implies that because of space-time dependent $(E,B)$, at every space position, there must be a thermal spread in velocities of charged particles at this space position.

The VE reflects the conservation in total particle number. If the general form $\alpha(r,t)\delta[v-\beta(r,t)]$ can meet a VE, this will mean that the total number of particles described by $\alpha(r,t)\delta[v-\beta(r,t)]$ is conserved and hence all particles are frozen to have a velocity equalling to $\beta(r,t)$. Obviously, above strict mathematical analysis denies this conjecture.

Because the total number of particles described by $\alpha(r,t)\delta[v-\beta(r,t)]$ is not conserved, or space-time dependent $(E,B)$ must require some particles being "evaporated" to be of $v \neq \beta(r,t)$, the distribution function of a realistic charged particles system (whose $(E,B)$ are space-time dependent) is therefore always of thermal spread. This implies that Eq.(6) always reflects the effect of the thermal spread. Especially, it is incorrect to put Eq.(6) in the zero-temperature limit because Eq.(6) is derived from a VE whose solution cannot be zero-temperature type $\alpha(r,t)\delta[v-\beta(r,t)]$. Namely, the equation

$$\partial_t \frac{u}{\sqrt{1-u^2}} + u \cdot \nabla \frac{u}{\sqrt{1-u^2}} + \frac{e}{m}[E + u \times B] = 0 \qquad (8)$$

is invalid.

Thus, charged particles must be described by a finite-temperature type distribution function $f$. However, for any finite-temperature type $f$, we can always construct a zero-temperature kernel $f_0$

$$f_0 = [\int f * \delta\left(v - \frac{\int v f d^3v}{\int f d^3v}\right) d^3v] * \delta\left(v - \frac{\int v f d^3v}{\int f d^3v}\right) \qquad (9)$$

which meets $\frac{\int v f d^3v}{\int f d^3v} = \frac{\int v f_0 d^3v}{\int f_0 d^3v}$. As above discussed, the subsystem described by $f_0$ will have particle exchange with another one described by $f - f_0$ and hence $f_0$ cannot satisfy the VE.

Because the $v$-dependence of $f_0$ is known, (i.e., a Dirac function), the equation of $f_0$ can be derived straightforward from following mathematical relations (where $n_0 = \int f_0 d^3v$)

$$\partial_t f_0 = \partial_t n_0 * \delta(v - u(r,t)) - n_0 * \partial_t u * \delta'; \qquad (10)$$

$$\nabla_r f_0 = \nabla_r n_0 * \delta(v - u(r,t)) - n_0 * \nabla_r u * \delta' \qquad (11)$$

and hence

$$[\partial_t + v \cdot \nabla_r - e[E(r,t) + v \times B(r,t)] \cdot \partial_p] f_0$$
$$= [\partial_t n_0 + v \cdot \nabla_r n_0] * \delta - n_0 [\partial_t u + v \cdot \nabla_r u] \delta' - \frac{e}{m}[E(r,t) + v \times B(r,t)]\left(\sqrt{1 - v \cdot v}\right)^3 \cdot \partial_v (n_0 \delta)$$
$$= [\partial_t n_0 + u \cdot \nabla_r n_0] * \delta - n_0 [\partial_t u + v \cdot \nabla_r u] \delta' - \frac{e}{m} n_0 [E(r,t) + v \times B(r,t)]\left(\sqrt{1 - v \cdot v}\right)^3 \cdot \delta'$$
$$= [\partial_t n_0 + u \cdot \nabla_r n_0] * \delta - n_0 [\partial_t u + v \cdot \nabla_r u] \delta' - \frac{e}{m} n_0 [E(r,t) + u \times B(r,t)]\left(\sqrt{1 - u \cdot u}\right)^3 \cdot \delta' \qquad (12)$$

Here, we have used mathematical relations

$$v \cdot \nabla_r n_0 * \delta(v - u(r,t)) = u \cdot \nabla_r n_0 * \delta(v - u(r,t)) \qquad (13)$$

$$\frac{\left(\sqrt{1 - v \cdot v}\right)^3 - \left(\sqrt{1 - u \cdot u}\right)^3}{v - u} * \delta(v - u(r,t)) = 0 \qquad (14)$$

Note that Eq.(12) is actually an integral-differential equation of a quantity $Q$ which is defined as

$$Q = [\partial_t + v \cdot \nabla_r - e\left[E(r,t) + v \times B(r,t)\right] \cdot \partial_p] f_0 + v \cdot \nabla_r \frac{\int v f_0 d^3 v}{\int f_0 d^3 v} \partial_v f_0 \qquad (15)$$

and hence can be re-written as

$$0 = Q - \left[\int Q * d^3 v\right] * \delta + \left[\int (Q * v) d^3 v\right] * \delta'. \qquad (16)$$

Clearly, a strict solution of Eq.(16) reads

$$0 = Q = [\partial_t + v \cdot \nabla_r - e\left[E(r,t) + v \times B(r,t)\right] \cdot \partial_p] f_0 + v \cdot \nabla_r \frac{\int v f_0 d^3 v}{\int f_0 d^3 v} \partial_v f_0 \qquad (17)$$

Compared with the VE or Eq.(1), there is a new operator $v \cdot \nabla_r u \partial_v$ appearing in Eq.(17). Due to this new operator, the continuity equation associated with $n_0$ becomes

$$\partial_t n_0 + u \cdot \nabla_r n_0 = 0, \qquad (18)$$

rather than our familiar

$$\partial_t n_0 + u \cdot \nabla_r n_0 = -n_0 \nabla_r \cdot u, \qquad (19)$$

or

$$\partial_t n_0 + \nabla_r \cdot (n_0 u) = 0. \qquad (20)$$

This new operator reflects the subsystem described by $f_0$ having particle exchange with other. Namely, because $E$ is space-time dependent, a charged particle system cannot be at zero-temperature state in which at any space position, all particles have a same velocity. Space-inhomogeneous $E$ will lead to, in some space positions, the temperature differing from 0 and hence thermal spread in particles' velocities appearing (which means some particles being out of the kernel group described by $f_0$ and into the hollow group described by $f - f_0$).

Actually, we can verify above conclusion via *reduction to absurdity*. If $n_0 * \delta(v - u(r,t))$ meets a VE, i.e. $[\partial_t + v \cdot \nabla_r - e\left[E(r,t) + v \times B(r,t)\right] \cdot \partial_p][n_0 * \delta(v - u(r,t))] = 0$, there will be a standard continuity equation, i.e. Eq.(20), and a standard fluid motion equation, i.e. Eq.(8). However, if directly appling the operator $[\partial_t + v \cdot \nabla_r - e\,[E(r,t) + v \times B(r,t)] \cdot \partial_p]$ on $n_0 * \delta(v - u(r,t))$, we will find that there exists

$$[\partial_t + v \cdot \nabla_r - e\,[E(r,t) + v \times B(r,t)] \cdot \partial_p]\,[n_0 * \delta(v - u(r,t))] = [n_0 \nabla_r \cdot u]\,\delta(v - u(r,t)),$$

which contradicts with the premise "$n_0 * \delta(v - u(r,t))$ meets a VE". This suggests that the statement "$n_0 * \delta(v - u(r,t))$ meets a VE" is not true. In contrast, $Q = 0$ does not lead to similar contradiction.

Likewise, following a standard procedure, we can obtain a macroscopic fluid motion equation from Eq.(17)

$$\partial_t \frac{u}{\sqrt{1 - u^2}} + \frac{e}{m}[E + u \times B] = 0 \qquad (21)$$

A certain relation between $u$ and $(E, B)$ is given by Eq.(21).

The VE can be written as

$$\begin{aligned} 0 &= \left[\partial_t + v \cdot \nabla - [E + v \times B] \cdot \partial_{p(v)}\right] f \\ &= \left[\partial_t - E \cdot \partial_{p(v)}\right] f + v \cdot \left[\nabla - B \times \partial_{p(v)}\right] f. \end{aligned} \qquad (22)$$

if $f$ is a strict solution of the VE, any mono-variable function of $f$, or $g(f)$, is also a strict solution.

For the case in which $E$ and $B$ are running waves of a phase velocity $\frac{1}{\eta}c$, i.e. $E = E(r - \frac{1}{\eta}ct)$ and $B = B(r - \frac{1}{\eta}ct)$, we should note a relation between $E$ and $B$: $E = -\frac{1}{\eta}c \times B + \nabla \Phi(r - \frac{1}{\eta}ct)$, which arises from $\nabla \times E = -\partial_t B$. Here, $\Phi(r - \frac{1}{\eta}ct)$ is a scalar function but cannot be simply taken as electrastatic potential (because $-\frac{1}{\eta}c \times B$ also has divergence or $\nabla \cdot \left(-\frac{1}{\eta}c \times B\right) = \frac{1}{\eta}c \cdot \nabla \times B \neq 0$). In this case, the VE can be further written as

$$\begin{aligned} 0 &= \left[\partial_t - E \cdot \partial_{p(v)}\right] f + v \cdot \left[\nabla - B \times \partial_{p(v)}\right] f \\ &= \left[\partial_t + \frac{1}{\eta}c \times B \cdot \partial_{p(v)}\right] f + v \cdot \left[\nabla - B \times \partial_{p(v)}\right] f - \nabla \Phi \cdot \partial_{p(v)} f \\ &= \left[\partial_t + \frac{1}{\eta}c \cdot B \times \partial_{p(v)}\right] f + v \cdot \left[\nabla - B \times \partial_{p(v)}\right] f - \nabla \Phi \cdot \partial_{p(v)} f \\ &= \left(v - \frac{1}{\eta}c\right) \cdot \left[\nabla - B \times \partial_{p(v)}\right] f - \nabla \Phi \cdot \partial_{p(v)} f \end{aligned} \qquad (23)$$

It is easy to verify that any function of $p + \int E(r - \frac{1}{\eta}ct)dt$ will meet

$$0 = \left[\partial_t - E \cdot \partial_{p(v)}\right] g \left(p + \int E(r - \frac{1}{\eta}ct)dt\right)$$
$$= [-\frac{1}{\eta}c \cdot \nabla + \frac{1}{\eta}c \cdot B \times \partial_{p(v)}]g$$
$$= -\frac{1}{\eta}c \cdot [\nabla - B \times \partial_{p(v)}]g, \qquad (24)$$

where we have used the property $\partial_t \int E(r - \frac{1}{\eta}ct)dt = -\frac{1}{\eta}c \cdot \nabla \int E(r - \frac{1}{\eta}ct)dt$. Thus, if $\nabla \Phi \equiv 0$, any mon-variable function of $p + \int E(r - \frac{1}{\eta}ct)dt$, or $g\left(p + \int E(r - \frac{1}{\eta}ct)dt\right)$, will be a strict solution of the VE. On the other hand, for more general $\nabla \Phi$, we can find that any mono-variable function of $\sqrt{1 + \frac{p^2}{c^2}} - \frac{1}{\eta}c \cdot p + \Phi$, or $g\left(\sqrt{1 + \frac{p^2}{c^2}} - \frac{1}{\eta}c \cdot p + \Phi\right)$, is a strict solution of the VE. According to Eq.(23), $\partial_p \left[\sqrt{1 + \frac{p^2}{c^2}} - \frac{1}{\eta}c \cdot p\right]$ will contribute a vector parallel to $\left(v - \frac{1}{\eta}c\right)$ and hence make the operater $\left(v - \frac{1}{\eta}c\right) \cdot B \times \partial_{p(v)}$ has zero contribution.

Therefore, for coherent self-consistent fields $E = E(r - \frac{1}{\eta}ct)$ and $B = B(r - \frac{1}{\eta}ct)$, the microscopic distribution $f$, if $\nabla \Phi \equiv 0$, can be described by a positive-valued function of $p + \int E(r - \frac{1}{\eta}ct)dt$, for example $\exp\left[-\left(p + \int E(r - \frac{1}{\eta}ct)dt\right)^2\right]$, $\sin^2(\exp\left[-\left(p + \int E(r - \frac{1}{\eta}ct)dt\right)^2\right])$, etc. We can further pick out reasonable solutions according to the definition of $u$

$$u = \frac{\int \frac{p}{\sqrt{1+p^2}} g\left(p + \int E(r - \frac{1}{\eta}ct)dt\right) d^3p}{\int g\left(p + \int E(r - \frac{1}{\eta}ct)dt\right) d^3p}. \qquad (25)$$

Likewise, same procedure exists for more general $\nabla \Phi$ and $g\left(\sqrt{1 + \frac{p^2}{c^2}} - \frac{1}{\eta}c \cdot p + \Phi\right)$.

Actually, a set of macroscopic functions $(E, B, u)$ can have multiple microscopic solutions of corresponding VE. Therefore, usually we know the microscopic distribution from its initial condition $f(r, p, t = 0)$. From the function dependence of $f(r, p, t = 0)$ on $p$, we can obtain the function dependence of $g$ on $\sqrt{1 + \frac{p^2}{c^2}} - \frac{1}{\eta}c \cdot p$ and hence determine detailed function form of $g$.

Detailed procedure of determining function form of $g$ is described as follows: We can seek for special space position $R$ in which $E(R, 0) = -\frac{1}{\eta}c \times B(R, 0)$, or $\nabla \Phi(r, 0)|_{r=R} = 0$, exists. The initial distribution at $R$, i.e., $f(R, p, t = 0)$, is thus a mono-variable function $p$. At the same time, two expressions are equivalent and hence there is $f(R, p, t = 0) =$ $g\left(K+\Phi\left(R,0\right)\right)=g\left(K\right)$. where $K=\sqrt{1+\frac{p^2}{c^2}}-\frac{1}{\eta}c\cdot p$ and $\Phi\left(R,0\right)=0$ (if $\nabla\Phi\left(r,0\right)|_{r=R}=0$). Because of certain relation between $p$ and $K$, once the expansion coefficients $c_i$ in $f\left(R,p,t=0\right)=\sum_i c_i p^i$ is known, the expansion coefficients $d_i$ in $g\left(K\right)=\sum_i d_i K^i$ is also esay to be calculated.

BGK modes are typical analytic strict solutions in $B\equiv 0$ case [19]. The VE in this case reads

$$0=\left[\partial_t - E\left(r-\frac{1}{\eta}ct\right)\cdot\partial_{p(v)}\right]f + v\cdot\nabla f, \qquad (26)$$

whose solutions are mono-variable functions of $\phi\left(r-\frac{1}{\eta}ct\right)+\sqrt{1+p^2}-\frac{1}{\eta}c\cdot p$, where $\phi\left(r-\frac{1}{\eta}ct\right)$ is scalar potential and $E=-\nabla\phi$. Moreover, there is a similar procedure of determining function form of $g$.

Here, we should note that $K$ is a nonlinear function of $p$ (or $v$) and the maximum value of $K$, or $K_{\max}$ is reached at $v=\frac{1}{\eta}c$. Thus, even $g$ is a Dirac function of $K+\Phi$, $g$ cannot be a Dirac function of $v$, i.e. $g\~\delta\left(v-u\left(r,t\right)\right)$. The nonlinear function relation between $K$ and $v$ determines that $g$ is at least a summation of two Dirac components: $g=f_1\left(r,t\right)\delta\left(v-u_1\left(r,t\right)\right)+f_2\left(r,t\right)\delta\left(v-u_2\left(r,t\right)\right)+....$ This agree with previous conclusion that functions of a general form $f_1\left(r,t\right)\delta\left(v-u_1\left(r,t\right)\right)$ cannot meet VE.

Moreover, we should also note that because of nonlinear function relation between $g$ and $K+\Phi$, the maximum of $g$, or $g_{\max}$, is usually reached at $K+\Phi\neq K_{\max}+\Phi$. Namely, if $g_{\max}$ is reached at $K=K_{g\max}$, this $K_{g\max}$ usually corresponds to two values of $v$. In contrast, $K_{\max}$ merely corresponds to a value of $v$. Thus, the contour plot of $g$ in the phase space will take on complicated structures, such as hole, island etc.

Above results have displayed a universal method of obtaining both exact macroscopic and microscopic information from the V-M system. The exact microscopic distribution is found to be an extension of the well-nown BGK modes. The exact closed equation set of macroscopic self-consistent fields is found to disentangle with thermal pressure.

This work is supported by 973 project in China.